\documentclass[twocolumn,superscriptaddress,nofootinbib,preprintnumbers,amsmath,amssymb,floatfix,showpacs]{revtex4-1}

\usepackage{graphicx}
\usepackage{amssymb}
\usepackage{dcolumn}
\usepackage{bm}
\usepackage{epsfig}
\usepackage{setspace}
\usepackage[caption=false]{subfig}
\usepackage{xcolor}
\usepackage{natbib}
\begin{document}

\title{Dynamical net-proton fluctuations near a QCD critical point}

\author{Christoph Herold}
\affiliation{School of Physics, Suranaree University of Technology, 111 University Avenue, Nakhon Ratchasima 30000, Thailand}

\author{Marlene Nahrgang}
\affiliation{Department of Physics, Duke University, Durham, North Carolina 27708-0305, USA}

\author{Yupeng Yan}
\affiliation{School of Physics, Suranaree University of Technology, 111 University Avenue, Nakhon Ratchasima 30000, Thailand}

\author{Chinorat Kobdaj}
\affiliation{School of Physics, Suranaree University of Technology, 111 University Avenue, Nakhon Ratchasima 30000, Thailand}

\email{herold@g.sut.ac.th}

\begin{abstract}

We investigate the evolution of the net-proton kurtosis and the kurtosis of the chiral order parameter near the critical point in the model of nonequilibrium chiral fluid dynamics. The order parameter is propagated explicitly and coupled to an expanding fluid of quarks and gluons in order to describe the dynamical situation in a heavy-ion collision. We study the critical region near the critical point on the crossover side. There are two sources of fluctuations: non-critical initial event-by-event fluctuations and critical fluctuations. These fluctuations can be distinguished by comparing a mean-field evolution of averaged thermodynamic quantities with the inclusion of fluctuations at the phase transition. We find that while the initial state fluctuations give rise to flat deviations from statistical fluctuations, critical fluctuations reveal a clear structure of the phase transition. The signals of the critical point in the net-proton and sigma field kurtosis are affected by the nonequilibrium dynamics and the inhomogeneity of the space-time evolution but develop clearly.  
\end{abstract}

\pacs{5.75.-q, 47.75.+f, 11.30.Qc, 24.60.Ky, 25.75.Nq}

\vspace{2pc}

\maketitle

\section{Introduction}

At large temperatures and densities, strongly-interacting matter is expected to change from a confined hadronic phase to a deconfined phase of quarks and gluons where chiral symmetry is restored. As the partition function of quantum chromodynamics (QCD) cannot be solved perturbatively in the region of the phase transition, we have to rely on other techniques to study the QCD phase diagram. Lattice QCD has successfully discovered the crossover nature of the transition \cite{Aoki:2006we,Borsanyi:2010bp} and established continuum-extrapolated results for the QCD equation of state, both at small baryochemical potential $\mu_{\rm B}$ \cite{Borsanyi:2013bia,Bazavov:2014pvz}. Due to the fermionic sign problem, standard lattice QCD methods become computationally ineffective in the regime of large densities. A couple of methods have been developed to extend the region of current lattice QCD calculations to finite baryochemical potential \cite{Fodor:2001au,deForcrand:2002ci,Endrodi:2011gv}, but quantitative conclusions cannot yet be drawn. A QCD critical point has been excluded up to values of $\mu_{\rm B}/T\lesssim1$. Another approach to the QCD phase diagram, which reproduces lattice results at vanishing baryochemical potential and can be extended to larger densities, comes from solving a coupled set of Dyson-Schwinger equations for the quark and gluon propagators \cite{Fischer:2014ata, Eichmann:2015kfa}. Here, a potential critical point is found to be located at $(T^c, \mu^c_{\rm q})=(115,168) \textrm{ MeV}$.

Ratios of cumulants of conserved quantities like the net-baryon number or net-electric charge are sensitive to a critical point  \cite{Asakawa:2000wh,Stephanov:1998dy,Stephanov:1999zu} signaling the singularity of thermodynamic quantities via their relation to susceptibilities \cite{Cheng:2008zh,Asakawa:2009aj,Gupta:2011wh,Friman:2011pf}. In this context higher-order cumulants are of special interest because they are more sensitive to the correlation length of fluctuations \cite{Stephanov:2008qz,Athanasiou:2010kw}. From universality arguments it has been demonstrated that the critical contributions to the kurtosis, in particular, may become negative approaching the critical point from the crossover side in heavy-ion collision experiments \cite{Stephanov:2011pb}.
This leads to the expectation of measuring a decreasing kurtosis lowering the beam energies, followed by a more complicated non-monotonic structure depending on the interplay of the location of the QCD critical point and freeze-out conditions for fluctuation measures.

Measurements of the net-proton and net-charge kurtosis and skewness have been reported by the STAR collaboration \cite{Aggarwal:2010wy,Adamczyk:2013dal}, where significant deviations from the hadron resonance gas and UrQMD calculations were found at lower beam energies. 

To understand the experimental data, it is important to develop dynamical models which are able to describe nonequilibrium effects of the QCD phase transition. Even if thermalization times are small during the evolution of the system created in a heavy-ion collision and local equilibrium is thus achieved, near the critical point the thermalization time diverges with a certain power of the correlation length given by the dynamical universality class \cite{HALPERIN:1969zza}. This phenomenon is called critical slowing down and limits the divergence of fluctuations due to finite-time effects. A phenomenological approach has been applied in \cite{Berdnikov:1999ph} to understand the growth of the correlation length in an evolving system. It was found that the correlation length does not grow beyond $1.5-2$~fm, but memory effects let the system remain correlated for a longer period than expected in equilibrium. Expanding Fokker-Planck dynamics in terms of powers of the correlation length over the system size in the scaling regime, the importance of memory effects was underlined in \cite{Mukherjee:2015swa}. In a real-time evolution of non-Gaussian moments the magnitude and the sign of the critical contributions could be significantly different from the equilibrium expectations. In this model a simplified and homogeneous expansion was assumed and the back-reaction of the order parameter fluctuations on the surrounding matter was not taken into account.

In this work, we focus on the real-time evolution of the fluctuations of the order parameter for chiral symmetry, the sigma field, as obtained in a coupled dynamics. While the sigma field is propagated explicitly via a stochastic-relaxational equation, it interacts with a fermionic heat bath which expands fluid dynamically \cite{Nahrgang:2011mg}. It has been demonstrated in \cite{Herold201414,Nahrgang:2011mv,Herold:2013bi,Herold:2014zoa}, that this model is able to describe critical slowing down as well as spinodal decomposition within a dynamical setup. 

We follow the evolution of the system over various hypersurfaces of constant energy density of the coupled system and compare the fluctuations in the sigma field to the fluctuations in net-proton numbers which are obtained from a Cooper-Frye particlization prescription. We furthermore give a comparison to a mean-field evolution in order to pin down the fluctuations stemming from the initial state versus dynamical fluctuations near the phase transition. 

We begin with a description of nonequilibrium chiral fluid dynamics (N$\chi$FD) in Sec.~\ref{sec:model}, including the equations of motion and a brief description of the implementation of initial state and particlization. In Sec.~\ref{sec:results}, we investigate the dynamics of the net-proton and sigma field kurtosis for an evolution in the crossover regime near the critical point. We conclude and give a brief outlook in Sec.~\ref{sec:summary}.

\section{Nonequilibrium chiral fluid dynamics}
\label{sec:model}

We study the fluid dynamical evolution in a heavy-ion collision using a quark-meson model with dilaton field, 
\begin{subequations}
\label{eq:Lagrangian}
\begin{align}
{\cal L}&=\overline{q}\left(i \gamma^\mu \partial_\mu-g_{\rm q} \sigma\right)q + \frac{1}{2}\left(\partial_\mu\sigma\right)^2 
+ \frac{1}{2}\left(\partial_\mu\chi\right)^2\\\nonumber 
        &\phantom{=} + {\cal L}_A- U_{\sigma}-U_{\chi}~, \\
\label{eq:LagrangianA}
 {\cal L}_A&=-\frac{1}{4}A_{\mu\nu}A^{\mu\nu}+\frac{1}{2}g_A^2\left(\frac{\chi}{\chi_0}\right)^2 A_\mu A^\mu~, \\
U_{\sigma}&=\frac{\lambda^2}{4}\left[\sigma^2-f_{\pi}^2\left(\frac{\chi}{\chi_0}\right)^2\right]^2-h\left(\frac{\chi}{\chi_0}\right)^2\sigma~, \\
U_{\chi}&=\frac{1}{4}B\left(\frac{\chi}{\chi_0}\right)^4\left[\ln\left(\frac{\chi}{\chi_0}\right)^4-1\right]~,
\end{align}
\end{subequations}
as introduced in \cite{Sasaki:2011sd}. In addition to the breaking and restoration of chiral symmetry it accounts for scale symmetry via the dilaton or glueball field $\chi$ that is identified with a gluon condensate. In the present version of this model we consider light quarks only, so $q=(u,d)$. The two condensates $\sigma$ and $\chi$ dynamically generate masses for the constituent quarks and gluons, thus allowing us to fix the coupling parameters $g_{\rm q}=3.37$ and $g_A=850 \textrm{ MeV}$ from the ground state nucleon and glueball masses. The additional parameters of the chiral sigma model are standard values: the pion decay constant of $f_\pi=93$~MeV, the pion mass $m_\pi=138$~MeV, the explicit symmetry breaking term $h=f_\pi m_\pi^2$ and the self-coupling $\lambda^2=\frac{m_\pi^2-m_\sigma^2}{2f_\pi^2}$. For more details, the reader is referred to \cite{Sasaki:2011sd,Herold:2014zoa}. 

In mean-field approximation, the effective thermodynamic potential reads 
\begin{equation}
\label{eq:effpot}
 V_{\rm eff}=\Omega_{q\bar q}+\Omega_{A}+U_{\sigma}+U_{\chi}+\Omega_0~,
\end{equation}
with the quark and gluon contributions
\begin{align} 
\Omega_{\rm q\bar q}&=-2 N_f N_c T\int\frac{\mathrm d^3 p}{(2\pi)^3} \left\{\ln\left[1+\mathrm e^{-\frac{E_{\rm q}-\mu}{T}}\right]\right.\\\nonumber 
                    &\phantom{-2 N_f N_c T\int\frac{\mathrm d^3 p}{(2\pi)^3}}+\left.\ln\left[1+\mathrm e^{-\frac{E_{\rm q}+\mu}{T}}\right]\right\}~, \\
\Omega_{A}&=2 (N_c^2-1) T\int\frac{\mathrm d^3 p}{(2\pi)^3} \left\{\ln\left[1-\mathrm e^{-\frac{E_A}{T}}\right]\right\}~,
\end{align}
which depend on temperature $T$ and quark chemical potential $\mu=\mu_{\rm B}/3$,
$\Omega_0$ in Eq.~(\ref{eq:effpot}) is an unimportant constant to set the total energy to zero in the ground state. The quasiparticle energies of constituent quarks and gluons are generated via their effective masses as $E_{\rm q}=\sqrt{p^2+m_{\rm q}^2}$ and $E_A=\sqrt{p^2+m_A^2}$.

The mean-field values of the condensate fields, $\langle\sigma\rangle$ and $\langle\chi\rangle$, are obtained by minimizing the effective thermodynamic potential $V_{\rm eff}$ via
\begin{equation}
\label{eq:equilibrium}
\left. \frac{\partial V_{\rm eff}}{\partial\sigma}\right |_{\sigma=\langle\sigma\rangle}=0~,~~\left. \frac{\partial V_{\rm eff}}{\partial\chi}\right |_{\chi=\langle\chi\rangle}=0~.
\end{equation}
In what we will call the mean-field evolution, the order parameter fields are set to their mean-field values neglecting fluctuations and the pressure is given by $p=-V_{\rm eff}$. The energy and quark number density of the system are thus evaluated as $e=T\partial p/\partial T +\mu n_{\rm q}-p$ and $n_{\rm q}=\partial p/\partial \mu$. This is equivalent to conventional deterministic fluid dynamical calculations using a chiral equation of state (EoS). 

From the curvature of the effective potential at the equilibrium value, the mass of the sigma field and thus the inverse correlation length are obtained as
\begin{equation}
 \label{eq:corrl}
 m_\sigma^2=\frac{1}{\xi_{\rm eq}^2}=\frac{\partial^2 V_{\rm eff}}{\partial\sigma^2}\bigg|_{\sigma=\langle\sigma\rangle}\, .
\end{equation}

In order to study nonequilibrium effects, we follow our previous works \cite{Nahrgang:2011mg,Nahrgang:2011vn,Herold:2013bi,Herold:2014zoa}, and propagate both order parameters explicitly. For the chiral condensate we derive a stochastic relaxation equation from the two-particle irreducible effective action as
\begin{equation}
\label{eq:eomsigma}
 \partial_\mu\partial^\mu\sigma+\eta_{\sigma}\partial_t \sigma+\frac{\delta V_{\rm eff}}{\delta\sigma}=\xi~,
\end{equation}
which takes into account interactions with the surrounding quark heat bath via a dissipative term and a stochastic noise field $\xi$. In the simplest approximation the noise is Gaussian

\begin{equation}
\label{eq:dissfluctsigma}
 \langle\xi(t,\vec x)\xi(t',\vec x')\rangle_\xi=\delta(\vec x-\vec x')\delta(t-t')m_\sigma\eta_{\sigma}\coth\left(\frac{m_\sigma}{2T}\right)~,
\end{equation}
 and has a vanishing expectation value $\langle\xi(t,\vec x)\rangle=0$. Due to the discretizing of the space-time delta function the noise term will be dependent on the lattice spacing \cite{CassolSeewald:2007ru}. In order to avoid this numerical cut-off dependence, we coarse-grain the noise term over the spatial extension of the equilibrium estimate for the correlation length as given in Eq. (\ref{eq:corrl}). 

The damping coefficient $\eta_{\sigma}$ depends on temperature and chemical potential,
\begin{equation}
\label{eq:dampingcoeff}
  \eta_{\sigma}=\frac{12 g^2}{\pi}\left[1-2n_{\rm F}\left(\frac{m_\sigma}{2}\right)\right]\frac{1}{m_\sigma^2}\left(\frac{m_\sigma^2}{4}-m_{\rm q}^2\right)^{3/2}~,
\end{equation}
 and vanishes near the critical point, where the mass of the sigma field becomes zero and the constituent quarks massive. 
Below the phase transition we use a damping coefficient of $\eta_{\sigma}=2.2/\textrm{fm}$ as has been estimated for the $\sigma-\pi$ interaction in \cite{Greiner:1996dx}.

In the temperature regime of interest the dilaton field only fluctuates minimally around its equilibrium value, as the restoration of scale symmetry occurs at much higher temperatures only. We, therefore, propagate small fluctuations according to the classical Euler-Lagrange equation of motion
\begin{equation}
\label{eq:eomchi}
 \partial_\mu\partial^\mu\chi+\frac{\delta V_{\rm eff}}{\delta\chi}=0~.
\end{equation}

The relaxation times of the constituent quarks and gluons are assumed to be much smaller than the long-wave length sigma mode and can thus be treated in local thermal equilibrium with the pressure

\begin{equation}
\label{eq:pressure}
 p(T,\mu; \sigma, \chi) = -\Omega_{q\bar q}-\Omega_{A}~.
\end{equation}

In order to conserve the total energy and momentum of the coupled system, the divergence of the energy-momentum tensor of the quark-gluon fluid $T^{\mu\nu}$ equals a source term from the sigma and dilaton fields

\begin{align}
\label{eq:fluidT}
\partial_\mu T^{\mu\nu}&=-\partial_\mu\left(T_\sigma^{\mu\nu}+T_\chi^{\mu\nu}\right)~,\\
\label{eq:fluidN}
\partial_\mu N_{\rm q}^{\mu}&=0~.
\end{align}
Thus, the fluid dynamical fields become stochastic as the evolution of the sigma field follows a stochastic differential equation. 

Recently, we have used this model to study the dynamical evolution through a first-order phase transition, where spinodal decomposition plays an important role \cite{Mishustin:1998eq,Sasaki:2007db,Randrup:2009gp,Randrup:2010ax,Steinheimer:2012gc}. We have demonstrated the formation of non-uniform structures in the energy and baryon density \cite{,Herold201414} and the dynamical enhancement of fluctuations in the medium \cite{Herold:2014zoa}. Such effects are especially interesting for upcoming experiments at FAIR \cite{Friman:2011zz} and NICA \cite{nica:whitepaper} which will make the region of high baryon densities in the phase diagram accessible.

\subsection{Initial state}

In this paper, we use event-by-event initial conditions, as opposed to previous publications, where we used an averaged initial state, usually a smooth sphere or ellipsoid. The initial energy and baryon density profiles are obtained from the UrQMD transport model \cite{Bass:1998ca,Bleicher:1999xi} run at a  center-of-mass energy per nucleon pair of $\sqrt{s_{\rm NN}}=19.7\textrm{ GeV/c}$. However, as the underlying EoS is different from that of the effective chiral model used here, we have to scale the resulting quantities such that we can investigate the region around the critical point during the fluid dynamical evolution.
With this set of initial conditions from UrQMD the fields are initialized at their local equilibrium values according to the temperature and baryochemical potential profiles. 
All events are generated with zero impact parameter.

The UrQMD initial state has been used in recent hybrid model calculations at lower beam energies \cite{Auvinen:2013sba} looking at observables like elliptic and triangular flow. It has been noted that it becomes less reliable at lower energies where different space-time regions might not thermalize along a contour of proper time, but gradually during the evolution. Also the impact of the core-corona separation is more important. We apply the UrQMD initial state on the crossover side to the right of the critical point, where the system traverses the critical region, but does not extend to extreme baryonic densities. Our study is of exploratory nature to investigate the effect of initial state fluctuations versus dynamical fluctuations near the critical point.

The second- and the fourth-order moments of the event-by-event volume averaged sigma-field fluctuations at initial proper time $\tau_0$ is $\langle\Delta\sigma^2\rangle_{0}=4\textrm{ MeV}^2$ and  $\langle\Delta\sigma^4\rangle_{0}=64\textrm{ MeV}^4$ respectively. The initial kurtosis is thus given by $\kappa\sigma^2=4\textrm{ MeV}^2$.

\subsection{Hypersurfaces of constant energy density and particlization}

In order to follow the evolution of the initial and the dynamical fluctuations we average the sigma field over hypersurfaces of constant energy density and look at the event-by-event fluctuations. The energy density is given by the sum of the local energy density of the fluid and of the order parameter fields
\begin{align}
\label{eq:edens}
e=&\,\,e_{\rm fluid}+\frac{1}{2}\left(\frac{\partial\sigma}{\partial t}\right)^2+\frac{1}{2}\left(\nabla\sigma\right)^2+U_{\sigma} \\\nonumber &+\frac{1}{2}\left(\frac{\partial\chi}{\partial t}\right)^2+\frac{1}{2}\left(\nabla\chi\right)^2+U_{\chi}~.
\end{align}
In order to make a first qualitative connection to experimental observables we apply a Cooper-Frye particlization prescriptions \cite{Cooper:1974qi,Cooper:1974mv} to produce protons and antiprotons from the fluid dynamical fields, in particular from the energy density in Eq. (\ref{eq:edens}) by the help of the Cornelius hypersurface finder developed in \cite{Huovinen:2012is}. Besides (anti-)protons we produce all non-strange particles implemented in the UrQMD model, such that the fully integrated energy, momentum, net-charge and net-baryon number are conserved exactly in each event. In general, the effect of sigma field fluctuations should couple to particle production, in particular (anti-)protons and pions, via an interaction term like $g\bar{p}\sigma p$ \cite{Stephanov:1999zu, Stephanov:2008qz, Athanasiou:2010kw}. These contributions will be considered in future work. In this study we do not apply a subsequent hadronic cascade, but work in this direction is underway.

\section{Results}
\label{sec:results}

\begin{figure}[t]
{
\centering
    \includegraphics[width=0.5\textwidth]{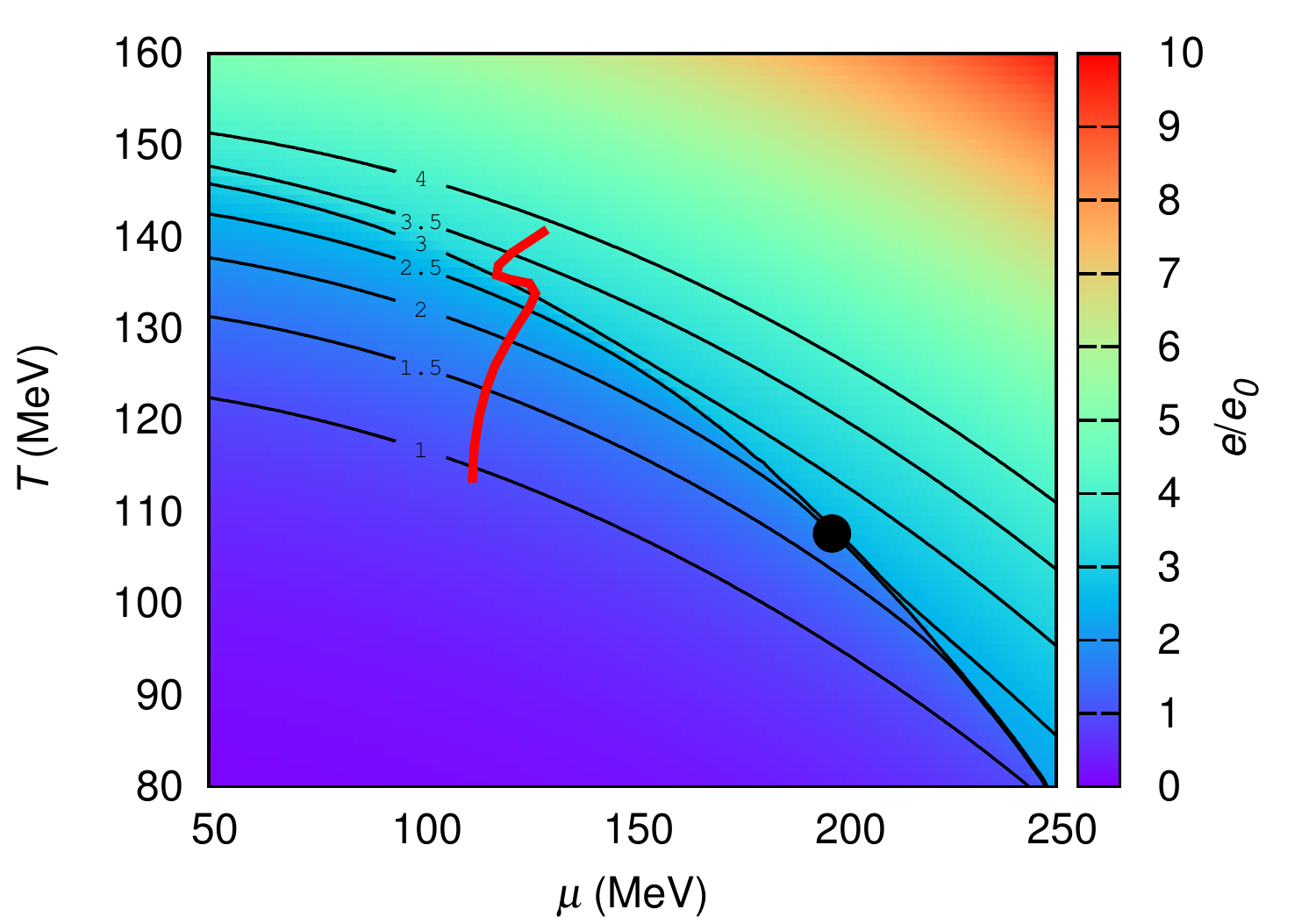}
}
\caption[trajectory]{(Color online) Event-averaged trajectory near the critical point (black dot). Lines of constant energy density are drawn to indicate the position of the particlization procedures.}
\label{fig:traj}
\end{figure}

The goal of this work is to study the real-time evolution of the kurtosis in a system which follows a trajectory on the crossover side near the critical point as seen in Fig.~\ref{fig:traj}. We have calculated the respective temperatures and quark chemical potentials as volume averages over different hypersurfaces of constant energy density and then averaged these values over a set of events. Along the phase boundary we notice a bending which typically occurs in the crossover fluid dynamical trajectories and isentropes of chiral effective models in mean-field approximation \cite{Kahara:2008yg,Nakano:2009ps,Herold201414}. We compare mean-field, where $\sigma=\langle\sigma\rangle$, $\chi=\langle\chi\rangle$, from Eq.~(\ref{eq:equilibrium}), and nonequilibrium evolutions, i.e. with the order parameter fields evolved according to Eqs.~(\ref{eq:eomsigma}, \ref{eq:eomchi}), of the system to disentangle initial fluctuations and dynamically evolved critical fluctuations. The hypersurface- and event-averaged quantities $T$ and $\mu$ for the trajectory do not differ significantly between the mean-field and the nonequilibrium evolution.

\begin{figure}[t]
{
\centering
    \includegraphics[width=0.5\textwidth]{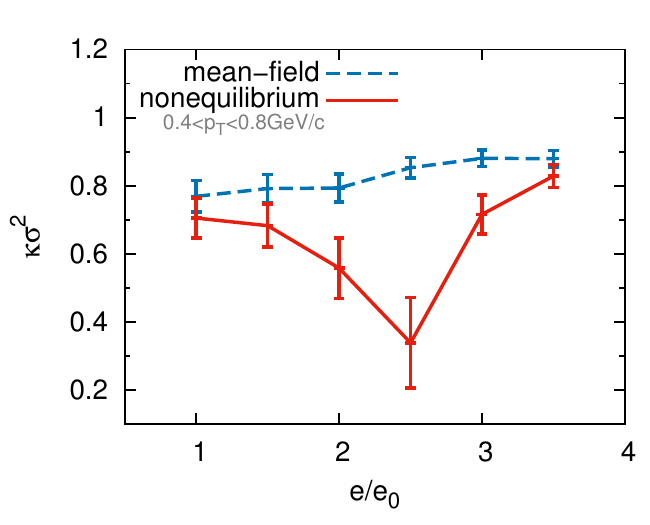}
}
\caption[kurtosis]{(Color online) Kurtosis of the net-proton number as function of freeze-out energy for mean-field and nonequilibrium evolution.}
\label{fig:kurt}
\end{figure}

\begin{figure}[t]
{
\centering
    \includegraphics[width=0.5\textwidth]{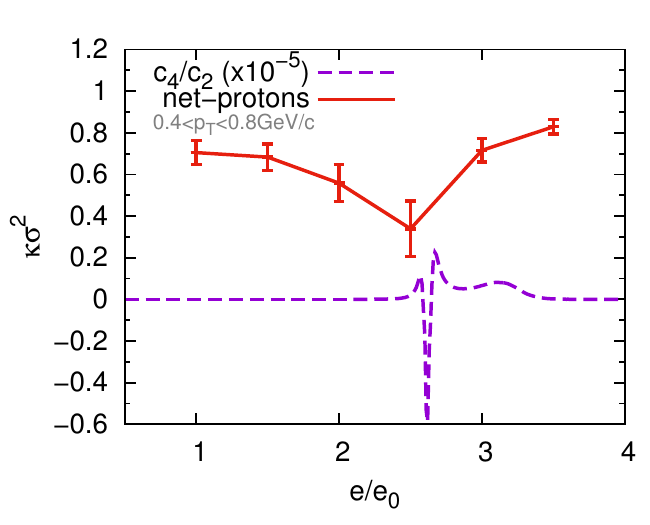}
}
\caption[kurtosis]{(Color online) Kurtosis of the net-proton number as function of particlization energy for nonequilibrium evolution compared with the thermodynamic equilibrium values calculated from generalized susceptibilities $c_4/c_2$.}
\label{fig:c4c2}
\end{figure}

First we extract the net-proton number kurtosis, a quantity that is also studied in experiment. It is calculated as
\begin{equation}
\kappa\sigma^2=\frac{\langle\Delta N^4\rangle}{\langle\Delta N^2\rangle}-3\langle\Delta N^2\rangle~,
\end{equation}
where $\Delta N= N-\langle N\rangle$ is the event-wise fluctuation in the net-proton number and $\langle\dots\rangle$ denotes an average over events.
For each event we extract the multiplicity with the above described particlization procedure over several hypersurfaces of constant energy density. Then, in order to observe fluctuations, we apply a cut in the rapidity of $|y|<0.5$ and in transverse momentum $0.4 \textrm{ GeV}<p_T<0.8 \textrm{ GeV}$. This is similar to the published data from the STAR Collaboration \cite{Aggarwal:2010wy,Adamczyk:2013dal}. The results are shown in Fig.~\ref{fig:kurt}, with a comparison between the mean-field and a nonequilibrium evolution. We can immediately see that for the mean-field scenario the values drop only slightly beyond unity, while in nonequilibrium we obtain a clear minimum at $e=2.5e_0$ where the kurtosis reaches a value of about $0.3$. Although there are event-by-event initial state fluctuations, they only result in  a flat behavior of the kurtosis as function of the particlization energy density. In the case of nonequilibrium, where we explicitly propagate the order parameter and allow for energy and momentum exchange between the field and the quark-gluon fluid, fluctuations build up when the systems passes the crossover region, resulting in a dip in the net-proton kurtosis. 
To compare this to the thermodynamic net-quark number kurtosis, we calculated the ratio of the generalized susceptibilities $c_4/c_2=\kappa\sigma^2$ along the trajectory in Fig.~\ref{fig:traj}. These susceptibilities are defined as \cite{Skokov:2010uh}
\begin{equation}
c_n=\frac{\partial^n (p/T^4)}{\partial(\mu_{\rm q}/T)^n}~.
\end{equation}
The result is shown in Fig.~\ref{fig:c4c2}, together with the nonequilibrium net-proton kurtosis. Note that in order to make a graphical comparison possible, we have scaled $c_4/c_2$ by a factor of $10^{-5}$. We see a sharp minimum with a negative quark-number kurtosis at $e=2.6e_0$, around the same point where we have the minimum in the net-proton kurtosis. We note two things: First, the equilibrium signal for criticality survives even in the dynamical environment representing the situation in a heavy-ion collision, though clearly less pronounced. Second, the resulting suppression of the net-proton kurtosis is spread out over a larger range of energy densities as a result of the inhomogenous medium and critical slowing down. It is important to remember that the values of $T$ and $\mu$ for the calculation of $c_4/c_2$ are averaged over the whole volume of the fireball. Therefore, a stronly negative kurtosis does not only occur at $2.6e_0$, but also around this value, whereas in smaller regions of space-time.

\begin{figure}[t]
{
\centering
    \includegraphics[width=0.5\textwidth]{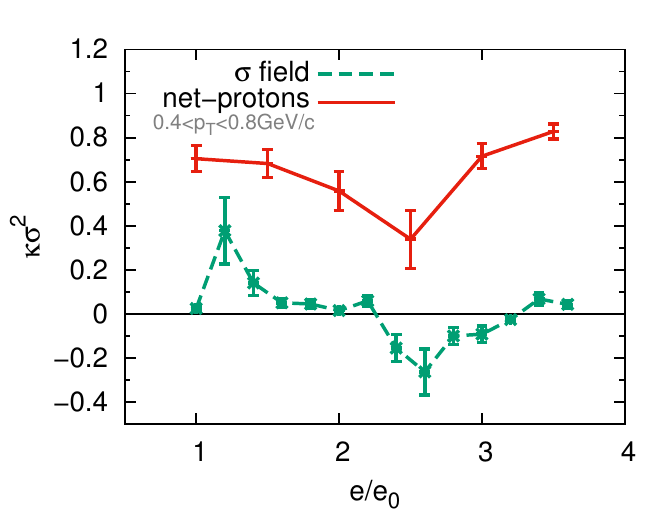}
}
\caption[proton]{(Color online) Net-proton kurtosis and kurtosis of the volume-averaged sigma field in comparison. The sigma field kurtosis is shown in unit $\textrm{MeV}^2$.}
\label{fig:sigma}
\end{figure}

In Fig.~\ref{fig:sigma} we compare the net-proton kurtosis of the nonequilibrium evolution to the kurtosis of the sigma field. We extract the latter one by volume-averaging the sigma field on hypersurfaces of constant energy density and subsequently calculating the event-by-event fluctuations of the obtained values $\langle\sigma\rangle_V$. Here we see a similar course in the two curves with a minimum at nearly the same energy density of about $2.5e_0$. 
In addition we observe a maximum of the sigma-field kurtosis at lower energy densities, which occurs below the equilibrium expectation for the phase transition and can be attributed to the importance of memory effects in a nonequilibrium evolution.

It is important to note that while the sigma field is not, the net-proton fluctuations are generally subject to global net-baryon number conservation \cite{Bleicher:2000ek,urqmdkurtosis,Bzdak:2012an}. At the presently investigated range of baryochemical potentials, however, baryon stopping should only have a negligible effect.

\section{Conclusions}
\label{sec:summary}

We have studied the net-proton kurtosis within the model of nonequilibrium chiral fluid dynamics, including a particlization procedure. This model captures the essential nonequilibrium dynamics of the order parameters at the QCD phase transition and critical point. We evaluated the kurtosis of both the sigma field and the net-proton number along an evolution near the critical point as a function of the energy density on a hypersurface. Here we compared mean-field fluid dynamical calculations to those with an explicit propagation of the order parameter fields, the chiral and gluon condensates. This takes into account the nonequilibrium evolution of the fluctuations via a stochastic relaxation equation. In the nonequilibrium case we found a minimum in both sigma and net-proton kurtosis at the same energy density. This minimum occurs around the same energy density as the minimum in the thermodynamic net-quark number kurtosis, implying that the suppression of the net-proton kurtosis is a remnant of the negative thermodynamic kurtosis. In comparison to that, a mean-field evolution without propagation of the order parameters shows a flat kurtosis as function of the energy density on the hypersurface.

The aim of future work will be to use a more realistic EoS, possibly including both hadronic and quark degrees of freedom. Models including quarks and hadrons have been studied in \cite{Dexheimer:2009hi,Steinheimer:2010ib,Turko:2013taa}, its parameters constrained by both lattice QCD data at small baryochemical potentials as well as neutron star properties at small temperatures. We will then study the kurtosis as a function of beam energy to compare results with the beam energy scan program at STAR. Finally, it is necessary to consider the evolution of produced fluctuations in a hadronic cascade to account for effects such as for example isospin randomization and charge diffusion processes \cite{Kitazawa:2011wh,Kitazawa:2012at,Nahrgang:2014fza,Sakaida:2014pya}.

\section*{Acknowledgements}

This work is funded by Suranaree University of Technology (SUT) and CHE-NRU (NV.10/2558) project. The authors thank Xiaofeng Luo and Volker Koch for fruitful discussions and Dirk Rischke for providing the SHASTA code that was used for the fluid dynamical simulation. The computing resources have been provided by the National e-Science Infrastructure Consortium of Thailand, the Center for Computer Services at SUT and the Frankfurt Center for Scientific Computing. MN acknowledges support from the U.S. Department of Energy under grant DE-FG02-05ER41367 and a fellowship within the Postdoc-Program of the German Academic Exchange Service (DAAD).

\section*{References}
\bibliographystyle{apsrev4-1}
\bibliography{mybib}

\end{document}